\documentclass{ws-procs9x6}
\usepackage{graphicx}

\newcommand{\fpi}{\mbox{$F_\pi$}}
\newcommand{\qsq}{\mbox{$Q^2$}}
\newcommand{\gevsq}{\mbox{(GeV/c)$^2$}}
\newcommand{\sigl}{\mbox{$\sigma_{\mathrm{L}}$}}
\newcommand{\sigt}{\mbox{$\sigma_{\mathrm{T}}$}}
\newcommand{\siglt}{\mbox{$\sigma_{\mathrm{LT}}$}}
\newcommand{\sigtt}{\mbox{$\sigma_{\mathrm{TT}}$}}

\begin{document}

\title{The Pion Form Factor}

\author{H.~P. BLOK}
\address{Department of Physics, Vrije Universiteit, Amsterdam, The Netherlands \\ 
E-mail: henkb@nat.vu.nl}

\author{G.~M. HUBER}
\address{Department of Physics, University of Regina, Regina SK, S4S 0A2 Canada\\ 
E-mail: huberg@uregina.ca}  

\author{D.~J. MACK}
\address{Physics Division, Jefferson Laboratory, Newport News, VA 23606, USA \\
E-mail: mack@jlab.org}

\maketitle

\abstracts{
The experimental situation with regard to  measurements of the pion charge
form factor is reviewed. Both existing data and planned experiments
are discussed.}

\section{Introduction}

The pion, and specifically its charge form factor, is of key interest
in the study of the quark-gluon structure of hadrons.
This is exemplified by the many calculations that treat the pion as
one of their prime examples.
One of the reasons is that the 
valence structure of the pion, being $\langle q\bar{q}\rangle$, is relatively 
simple.  Hence it is expected that the value of the four-momentum transfer 
squared \qsq , down to which a pQCD approach to the pion structure can be 
applied, is lower than for the nucleon.
Whereas, e.g., the proton form factors seem to be completely
dominated by constituent quark properties \cite{mil02} up to at least
\qsq\ = 10 - 20 \gevsq , recent estimates \cite{sch01} suggest that pQCD
contributions start to dominate
the pion form factor at \qsq\ $\geq$ 5 \gevsq .
Furthermore, the asymptotic normalization of the pion wave function,
in contrast to that of the nucleon, is known from the pion decay.
Within perturbative QCD one can then derive \cite{far79} 
\begin{equation}
\label{eq:asympt}
\lim_{Q^2\to\infty} F_{\pi} = \frac{8 \pi \alpha_s f_{\pi}^2} {Q^2},
\end{equation}
where $f_{\pi}$ is the pion decay constant.  The question is down
to which finite value of \qsq\ this relation is valid.
Thus the interest is in the transition from the soft regime, governed by all
kinds of quark-gluon correlations, at low \qsq , to the perturbative (including
next-to-leading order and transverse corrections) regime at high \qsq .

The charge form factor of the pion at very low values of \qsq ,
which is governed by the charge radius of the pion,
has been determined \cite{ame86} up to \qsq =0.28 \gevsq\ from
scattering high-energy pions from atomic electrons.
For the determination of the pion form factor at higher values of
\qsq\ one has to use high-energy electroproduction of pions on a nucleon, i.e.,
employ the $^1$H$(e,e'\pi^+)n$ reaction. For selected kinematical conditions 
this process can be described as quasi-elastic scattering of the electron from 
a virtual pion in the proton.
The cross section for this process can be written as
\begin{equation}
\label{eq:sigma1}
\frac{d^3 \sigma}{dE' d\Omega_{e'} d\Omega_\pi} = \Gamma_V
\frac{d^2 \sigma}{dt d\phi},
\end{equation}
where $\Gamma_V$ is the virtual photon flux factor, $\phi$ is the azimuthal 
angle of the outgoing pion with respect to the electron scattering plane and 
$t$ is the Mandelstam variable $t=(p_\pi-q)^2$. The two-fold differential cross
section can be written as
\begin{eqnarray}
\label{eq:sepsig}
2\pi \frac{d^2 \sigma}{dt d\phi} & = & 
   \epsilon \hspace{0.5mm} \frac{d\sigma_{\mathrm{L}}}{dt} +
   \frac{d\sigma_{\mathrm{T}}}{dt} + \sqrt{2\epsilon (\epsilon +1)}
   \hspace{1mm}\frac{d\sigma_{\mathrm{LT}}}{dt}
   \cos{\phi}  \nonumber \\
   & & + \epsilon \hspace{0.5mm}
   \frac{d\sigma_{\mathrm{TT}}}{dt} \hspace{0.5mm} \cos{2 \phi} ,
 \end{eqnarray}
 where $\epsilon$ is the virtual-photon polarization parameter.
The cross sections $\sigma_X \equiv \frac{d\sigma_{\mathrm{X}}}{dt}$ depend on
$W$, \qsq\ and $t$.
In the t-pole approximation the longitudinal 
cross section \sigl\ is proportional to the square of the pion form factor:
\begin{equation}
\label{eq:tpole_pi}
\sigma_L \propto   \frac { -t\,Q^2 } {(t-m_\pi^2)^2} F_{\pi}^2 .
\end{equation}
The $\phi$ acceptance of the experiment should be large enough for the
interference terms \siglt\ and \sigtt\ to be determined. Then, by taking
data at two energies at every \qsq , \sigl\ can be separated from \sigt\ 
by means of a Rosenbluth separation.

\section{Existing measurements}

The pion form factor has been studied for \qsq\ values from
0.4 to 9.8 \gevsq\ at CEA/Cornell \cite{beb78}.
For \qsq\ above 1.6 \gevsq\ these are at present still the only existing data.
In these experiments only in a few cases was an L/T separation performed,
and even then the resulting uncertainties in \sigl\ were so large that the L/T
separated data were not used. Instead, for the actual 
determination of the pion form factor, \sigl\ was calculated
by subtracting from the measured (differential) cross section a \sigt\ that
was assumed to be proportional to the total virtual photon cross section,
and no uncertainty in \sigt\ was included in this subtraction.
This means that the published values of \fpi\  have large additional model
uncertainties on top of the already relatively large statistical 
(and systematic) uncertainties.

\begin{figure}[t]
    \centerline{\includegraphics[width=3.6in]{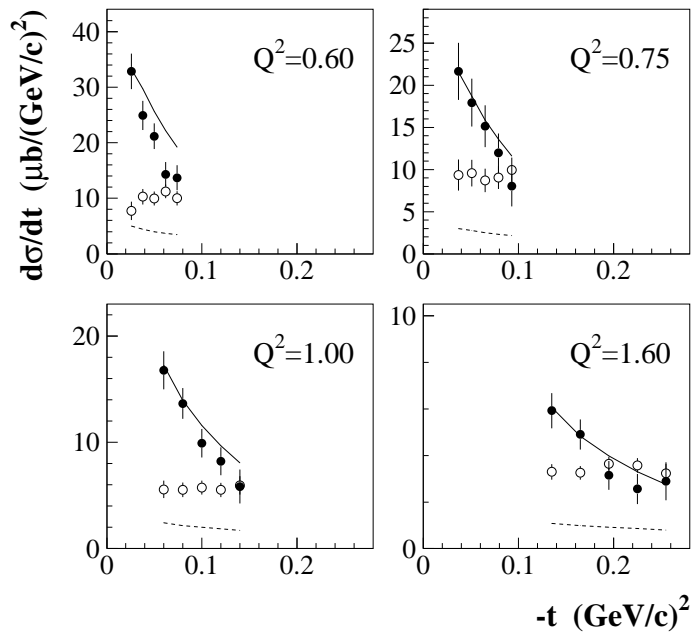}}
\caption{
        Separated cross sections \sigl\ and \sigt\ (full and open symbols,
        resp.) compared to the Regge model (full curve for L,  dashed curve for 
        T). The $Q^2$ values are in units of \gevsq .
        \label{fig:cross-sections}
}
\end{figure}
The pion form factor was also studied at DESY \cite{bra77} for
\qsq\ = 0.7 \gevsq .
In this case a full separation of all structure functions was performed.
We will come back to these data.

Recently the pion form factor was studied  \cite{vol01} at CEBAF for
\qsq\ = 0.6 - 1.6 \gevsq .
Using the High Momentum Spectrometer and the Short
Orbit Spectrometer of Hall C and 
electron energies between 2.4 and 4.0 GeV, data for the reaction 
$^1$H$(e,e'\pi^+)n$ were taken for central values of \qsq\ of 0.6, 0.75, 
1.0 and 1.6 \gevsq , at a central value of the invariant mass $W$ of 1.95 GeV.
Because of the excellent properties of the electron beam and experimental setup,
L/T separated cross sections could be determined with high accuracy.

The extracted cross sections are displayed in Figure \ref{fig:cross-sections}.
The error bars represent the combined statistical and systematic uncertainties.
As a result of the Rosenbluth separation the total error bars on
\sigl\ are enlarged considerably, resulting in typical error
bars of about 10\%. 

In order to determine the value of \fpi , the experimental data were
compared to the results of a Regge model by 
Vanderhaeghen, Guidal and Laget (VGL) \cite{van97}. In this model the pion 
electroproduction process is described as the exchange of Regge trajectories 
for $\pi$ and $\rho$ like particles.
The VGL model is compared to the data in Figure \ref{fig:cross-sections}.
Here the value of \fpi , which is a parameter in the model, was adjusted at
every \qsq\ to reproduce the \sigl\ data at the lowest value of $-t$.
The transverse cross section \sigt\ is underestimated, which can possibly
be attributed to resonance contributions at $W=1.95$ GeV that are not
included in the Regge model.
\newline
The t-pole dominance for \sigl\ at small $-t$ was checked by studying the
reactions $^2$H$(e,e'\pi^+)nn$ and $^2$H$(e,e'\pi^-)pp$, which gave within
the uncertainties a ratio of unity for the longitudinal cross sections.

\begin{figure}[t]
    \centerline{\includegraphics[angle=90,width=4.0in]{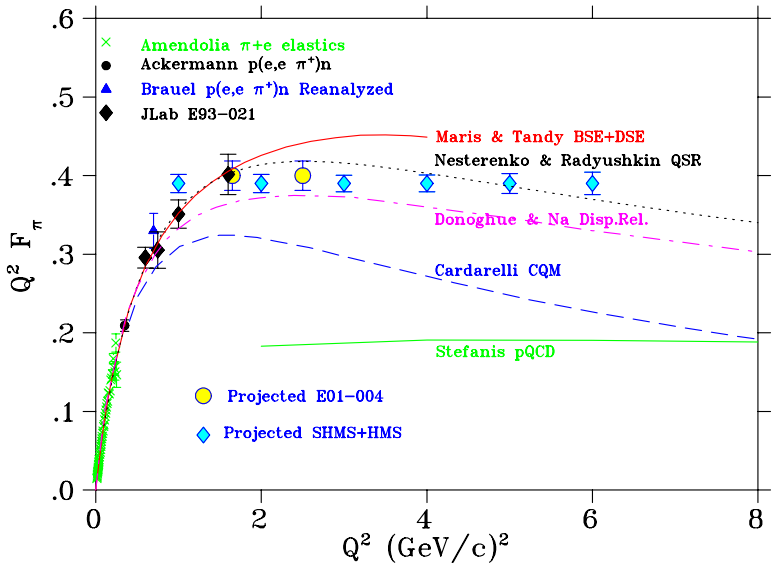}}
\caption{
        Existing and expected values for $F_\pi$
        in comparison to the results of several calculations.
        The model uncertainty is estimated to be about 5\%.
         The (model-independent) data from Ref. \protect\cite{ame86}
        are also shown.
        A monopole behaviour of the form factor obeying the measured
        charge radius is almost identical to the Maris and Tandy curve.
        \label{fig:piff}
}
\end{figure}

The comparison with the \sigl\ data shows that the $t$ dependence in the VGL 
model is less steep than that of the experimental data. 
As suggested by the analysis \cite{gut72} of older data, where a similar
behaviour was observed, we attributed this discrepancy to the presence of
a  small negative background contribution to the longitudinal cross
section, presumably again due to resonances.
The values of \fpi , extracted taking this into account, are shown
in Figure~\ref{fig:piff}.

For consistency we have determined \fpi\ in the same way from the cross
sections at \qsq\ = 0.7 \gevsq , $W=2.19$ GeV from DESY \cite{bra77}.
The background term in \sigl\ was found to be smaller than in the
Jefferson Lab data, presumably because of the larger value of $W$.
The resulting best value for \fpi , also shown in 
Figure~\ref{fig:piff}, is larger by 12\% than the original result, which was 
obtained by using the Born term model by Gutbrod and Kramer \cite{gut72}.
Those authors used a phenomenological $t$-dependent function,
whereas the Regge model by itself gives a good description of the
$t$-dependence  of the (unseparated) data from Ref. \cite{beb78}.

The data for \fpi\ in the region of \qsq\ up to 1.6 \gevsq\ globally follow
a monopole form obeying the pion charge radius \cite{ame86}.
It should be mentioned that the older Bebek 
data in this region suggested lower \fpi\ values.  However, as mentioned, they 
did not use L/T separated cross sections, but took a prescription for 
\sigt . Our measured data for \sigt\ indicate that the values used 
were too high, so that their values for \fpi\ came out systematically low.

In Figure~\ref{fig:piff} the data are also compared to a sample of theoretical
calculations. The model by Maris and Tandy \cite{mar00} provides a good 
description of the data. It is based on the Bethe-Salpeter equation with 
dressed quark and gluon propagators, and includes parameters that were 
determined without the use of \fpi\ data. The data are also well described by
the QCD sum rule plus hard scattering estimate of Ref.~\cite{nes82}. Other
models \cite{car94,don97} were fitted to the older \fpi\ data and therefore 
underestimate the present data. Figure \ref{fig:piff} also includes the results
from a perturbative QCD calculation \cite{ste99}.
Apart from the basic dependence given by Eq. 1, but extended to
next-to-leading order, it includes transverse
momenta of the quarks, Sudakov factors, and a way to regularize the
infrared divergence. As a result the value of \qsq \fpi\ is about
constant at 0.18 over the whole range of \qsq\ shown. Other pQCD
calculations yield similar results, but with a lower value
of \qsq \fpi \cite{jak93}.
Hence it is clear that in the region below \qsq $\approx 2$ \gevsq ,
where accurate data exist, soft contributions are much larger than
pQCD ones.
For this reason it is highly interesting to get reliable data at
higher values of \qsq .

\section{Future experiments}

The JLab experiment will be extended in the year 2003. Data will be taken at
\qsq\ = 2.5 \gevsq , the highest value compatible with the present
set-up, which is determined by the combination of the maximum momentum
of the SOS spectrometer (1.75 GeV/c) and the minium angle of the HMS
spectrometer (10.5 degrees).  The value of $W$ will be 2.20 GeV.
The increased value  of $W$ gives a smaller value of $-t_{min}$,
closer to the pole, and is in a region where the Regge
model is supposed to be more reliable.
Data will also be taken at \qsq\ = 1.6 \gevsq\ with $W=2.20$ GeV.
By comparing those to the existing ones, taken at $W=1.95$ GeV,
the model dependence in the extraction of \fpi\ will be gauged.

\begin{figure}[t]
    \centerline{\includegraphics[width=3.0in]{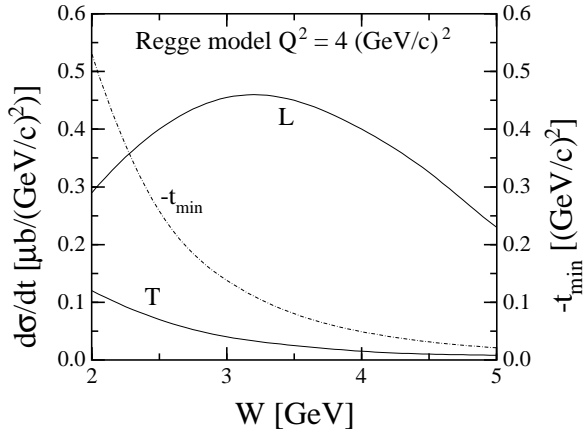}}
\caption{
        Dependence of the values of $-t_{min}$ and of \sigl\ and \sigt\
	at $t_{min}$, calculated with the Regge model, on $W$.
        \label{fig:wdep}
}
\end{figure}
With the planned upgrade of CEBAF to 12 GeV the pion form factor can
be studied \cite{12GeV} up to \qsq $\approx 6$ \gevsq\ with $W\geq 3.0$ GeV.
The HMS spectrometer will now be used to detect the scattered electron,
while the pion will be detected in the proposed SHMS spectrometer. Since the
direction of $\vec{q}$ is rather forward at high \qsq\ and $W$, the
small angle capability of SHMS is essential.
Taking advantage of the higher incoming energy, the value of $W$ can be
increased, with even more of the benefits described above. As a result,
contrary to common belief, the ratio of \sigl\ over \sigt\ will still
be favourable. This is illustrated in Figure~\ref{fig:wdep}. The value
of \sigt\ decreases with $W$ due to kinematical factors, but for \sigl\
this reduction is more than compensated for by the value of $-t_{min}$
getting smaller, i.e., closer to the pole. 

Figure~\ref{fig:piff} shows what data will be obtained, with the\
expected experimental accuracy. The model uncertainty is estimated
to be about 5\%.

\begin{table}[h]
\tbl{\normalsize
Kinematics for studying the kaon form factor at \qsq\ = 2.0 \gevsq\ 
and $W=3.3$ GeV, which gives $-t_{min}=0.120$ \gevsq .}
{\normalsize
\begin{tabular}{ccccc}
 $E_e$ (GeV)& $\theta_{e'}$ & $E_{e'}$ (GeV)  & $\theta_q$ & $\epsilon$ \\
 \hline 
 11.0 & 11.4 & 4.60 & 7.98 & 0.700 \\
  8.0 & 22.8 & 1.60 & 5.43 & 0.364 \\
 \hline 
\end{tabular}
\label{tab:kin_ka}
}
\end{table}
With the 12 GeV upgrade one can also start to think about studying
the kaon form factor. The formula for the Born  cross section in this case
is
\begin{equation}
\label{eq:tpole_ka}
\sigma_L \propto 
\frac { [-t+(M_\Lambda-M_p)^2]\,Q^2 } {(t-m_K^2)^2}\, F_K^2 .
\end{equation}
Clearly, accessible values of $-t$ will be much further from the pole.
However, by using large values of W one can hope that meaningful results
can be obtained up to \qsq\ $\approx$ 2 \gevsq .  Possible kinematics
at this \qsq\ are given in table \ref{tab:kin_ka}.
In the analysis one has also to take into account two-step processes
like forming first a $K^*$ particle, which then decays into a $K$.


\begin{thebibliography}{0}

\bibitem{mil02} G.~Miller, this workshop
\bibitem{sch01} W.~Schweiger,
	 {\it Nucl.~Phys.~Proc. Suppl.} {\bf 108}, 242 (2002)
\bibitem{far79} G.R. Farrar, D.R. Jackson,
	{\it Phys. Rev. Lett.} {\bf 43}, 246 (1979)
\bibitem{ame86} S.~R.~Amendolia {\em et al.},
	{\it Nucl.~Phys.} {\bf B277}, 168 (1986)
\bibitem{beb78} C.~J.~Bebek {\em et al.},
	{\it Phys.~Rev.} {\bf D17}, 1693 (1978)
\bibitem{bra77} P.~Brauel   {\em et al.},
	{\it Z.~Phys.} {\bf C3}, 101 (1979)
\bibitem{vol01} J.~Volmer {\em et al.},
	{\it Phys.~Rev.~Lett.} {\bf 86}, 1713 (2001);
J.~Volmer, PhD thesis, Vrije Universiteit, Amsterdam (2000), unpublished
\bibitem{van97} M.~Vanderhaeghen, M.~Guidal and J.-M.~Laget,
       {\it Phys.~Rev.} {\bf C57}, 1454 (1998);
       {\it Nucl.~Phys.} {\bf A627}, 645 (1997)
\bibitem{gut72} F.~Gutbrod and G.~Kramer,
	{\it Nucl.~Phys.} {\bf B49}, 461 (1972)
\bibitem{mar00} P.~Maris and P.~C.~Tandy,
	{\it Phys.~Rev.} {\bf C62}, 055204 (2000)
\bibitem{nes82} V.~A.~Nesterenko and A.~V.~Radyushkin,
	{\it Phys. Lett.} {\bf B115}, 410 (1982)
\bibitem{car94} F.~Cardarelli {\em et al.},
	{\it Phys.~Lett.} {\bf B332}, 1 (1994);
        {\it Phys.~Lett.} {\bf B357}, 267 (1995)
\bibitem{don97} J.F.~Donoghue and E.S.~Na,
	{\it Phys.~Rev.} {\bf D56}, 7073 (1997)
\bibitem{ste99} N.G.~Stefanis, W.~Schroers and H.-Ch.~Kim,
	{\it Phys. Lett.}  {\bf B449}, 299 (1999); 
	{\it Eur.~Phys.~J.} {\bf C18}, 137 (2000)
\bibitem{jak93} R.~Jakob and P.~Kroll,
	{\it Phys.~Lett.} {\bf B315}, 463 (1993) and {\bf B319}, 545 (1993);
	{\it J.~Phys.~G.} {\bf 22}, 45 (1996)
\bibitem{12GeV} "The Science Driving the 12 GeV Upgrade of CEBAF",
{\it Jefferson Lab Report}, Febr. 2001


\end{thebibliography}
\end{document}